\documentclass[a4paper,fleqn]{cas-dc}

\usepackage[numbers]{natbib}
\usepackage{subcaption}
\usepackage{placeins}

\def\tsc#1{\csdef{#1}{\textsc{\lowercase{#1}}\xspace}}
\tsc{WGM}
\tsc{QE}
\tsc{EP}
\tsc{PMS}
\tsc{BEC}
\tsc{DE}

\usepackage{xcolor}
\definecolor{darkred}{RGB}{210,60,60}
\definecolor{darkgreen}{RGB}{40,160,90}

\begin{document}
\let\WriteBookmarks\relax
\def\floatpagepagefraction{1}
\def\textpagefraction{.001}

\shorttitle{Measuring and Modelling Lag in a-Si Flat-Panel Detectors}
\shortauthors{Y. Huang et al.}

\title[mode=title]{Measuring and Modelling Lag in Amorphous Silicon Flat-Panel X-ray Detectors}

\author[1]{Yiyue Huang}

\author[1]{Benjamin Young}

\author[1]{Andrew Kingston}

\author[1]{Adrian Sheppard}
\cormark[1]
\ead{adrian.sheppard@anu.edu.au}

\affiliation[1]{
    organization={Department of Materials Physics, Research School of Physics, The Australian National University},
    city={Canberra},
    postcode={2600},
    state={ACT},
    country={Australia}
}

\cortext[cor2]{Corresponding author.}

\begin{abstract}
Detector lag, also referred to as afterglow, is a source of image degradation in flat-panel x-ray detectors, producing temporal artefacts that reduce image quality and quantitative accuracy. In this work, we present a robust and repeatable experimental framework for characterising long-term detector lag (from minutes to hours) under controlled step-up and step-down exposure transitions of particular relevance to tomography. The measured transition curves from an amorphous silicon detector with a CsI:Ti scintillator exhibit unexpected behaviour, with the detector response temporarily overshooting its final equilibrium intensity following step-up transitions, rather than exhibiting the gradual monotonic rise and decay typically expected. These curves were well fitted by multi-exponential functions, providing initial estimates for the depth of the charge traps in the scintillator. These parameters were incorporated into a proposed multi-trap rate equation model that reproduces the overall observed behaviour of both step-up and step-down transitions, including the overshoot that is not captured by previous models. Although discrepancies remain between the model and observation, this work establishes a reproducible methodology for detector lag characterisation and presents an improved physical model that offers greater insight into the mechanisms governing detector lag.
\end{abstract}


\maketitle

\section{Introduction}

Detector lag, also referred to as a memory effect, hysteresis, or afterglow, in inorganic scintillation crystals remains a persistent challenge in the design and operation of ionizing radiation detectors \cite{hsieh1997analysis,siewerdsen1999ghost}. It can be particularly problematic for quantitative applications of x-ray radiography and computed tomography (XCT) \cite{sato2015evaluation}. Following a change in incident x-ray flux, the emitted light from a scintillator does not immediately reach a new equilibrium. Instead, the signal exhibits a time-dependent relaxation, which can manifest as a slow rise or decay depending on whether the incident intensity increases or decreases. This temporal persistence introduces memory effects in the detector response, which can degrade image fidelity when acquisition conditions vary over time \cite{tanaka2010investigation}.


Beyond image artefacts, detector lag fundamentally limits the accuracy of quantitative imaging.  If lag effects could be fully understood and corrected, it would enable more reliable material decomposition, more accurate x-ray flux measurements, improved dynamic imaging, and higher precision in applications such as dual-energy XCT \cite{alvarez1976energy} and time-resolved studies \cite{chen2008prior}.

The behaviour of scintillators is governed by the populations of charge carriers within the material, specifically electrons and holes in the conduction and valence bands, as well as carriers captured in a distribution of trapping states introduced by dopants and defects in the semiconductor lattice \cite{bartram2006suppression}. In flat-panel x-ray detectors, each absorbed x-ray photon generates thousands of electron–hole pairs within the scintillator, which causes the emission of visible scintillation light. The trapping and subsequent recombination of these charge carriers, governs the detector’s temporal response, and gives rise to detector lag. These traps can capture and release carriers over a wide range of timescales, from microseconds to minutes or longer, thereby modulating the scintillation output. The resulting dynamics of the light production are determined by the depth, density, and recombination pathways associated with these trapping centres \cite{moretti2016deep}. For example, the CsI:Tl scintillator, which is widely used as XCT detectors, contains as many as twelve distinct trap types, including shallow and deep electron and hole traps associated with thallium activators, intrinsic defects, and impurity centres \cite{gridin2015kinetic}. The wide distribution of trap energies and recombination pathways gives rise to the complex temporal behaviour observed in detector lag.

The depth and prevalence of relevant traps are commonly characterised using thermoluminescence glow curves, obtained by progressively heating the scintillator after irradiation and recording the emitted light as trapped carriers are thermally released. Such measurements provide insight into the trap energy distribution and associated lifetimes, which are central to understanding lag behaviour and afterglow characteristics \cite{moretti2014radioluminescence}.

In many common inorganic scintillators, lag effects (lasting longer than a few seconds) are relatively small, typically at the level of a few percent, and are therefore often neglected in standard x-ray imaging and tomography workflows. However, even small temporal deviations can become significant in quantitative imaging applications. In particular, dual-energy decomposition methods \cite{alvarez1976energy} rely on accurate modelling of intensity changes across exposures, as do high-contrast industrial XCT applications, such as the quality inspection of additively manufactured parts. Small systematic errors in detector response can propagate through the reconstruction and decomposition processes, leading to larger errors in the estimated atomic number and density, visible artefacts, and bias in the recovered material properties.

Cesium iodide doped with thallium (CsI:Tl) is one of the most widely used scintillator materials in flat-panel detectors due to its high light yield, good stopping power, and the ability to grow the crystal in a columnar microstructure \cite{bartram1996efficiency}. This columnar morphology acts as a waveguide, reduces lateral light spread and therefore minimises pixel cross-talk, making CsI:Tl especially well suited for high-resolution x-ray microscopy and computed tomography \cite{nagarkar2004high, brecher2006suppression}. As a result, CsI:Tl is used almost exclusively in many laboratory-based XCT systems. However, despite these favourable properties, afterglow in CsI:Tl is particularly pronounced on timescales ranging from milliseconds to minutes, memory effects associated with deep traps have also been reported on much longer timescales, extending to tens of minutes or longer \cite{alfassi2009elimination, moretti2016deep}. These long-lived processes make lag a critical consideration in applications involving changing exposure conditions. Although these timescales are much longer than those typically encountered in medical XCT acquisitions, they become especially important in laboratory-based XCT for materials science, where individual scans commonly extend over several hours and the cumulative effects of detector lag can significantly affect quantitative image quality.


Previous work has developed both physics-based and empirical models to describe detector lag. Physics-based models are generally formulated as coupled rate equations describing the populations of free and trapped charge carriers and the radiative and non-radiative recombination processes governing scintillation light production \cite{gridin2015kinetic}. The release of trapped carriers is thermally activated, with the characteristic trap lifetime closely related to the trap depth through an Arrhenius (Boltzmann) relationship, such that deeper traps exhibit longer decay times. Bartram et al. \cite{bartram2006suppression} and Moretti et al. \cite{moretti2016deep} introduced coupled rate equations to describe carrier trapping and recombination dynamics in CsI:Ti and related crystals, including deep trapping states that can suppress afterglow. In parallel, Starman et al. \cite{starman2012nonlinear} proposed multi-exponential impulse response models as a practical approach for modelling and correcting for lag in imaging systems, and demonstrated improvements in XCT scans. These approaches are discussed in more detail in Sec.~\ref{sec:background}.

In this work, we experimentally probe detector lag under controlled intensity transitions and analyse the resulting dynamics using both empirical and physically motivated models. While previous studies have primarily focused on memory effects and afterglow under simplified or isolated exposure conditions (transitions to and from zero to exposure), this work is, to the best of our knowledge, the first to systematically investigate exposure changes typical of those encountered during an XCT scan. It is also the first to examine detector lag over the experimentally relevant timescales of laboratory XCT acquisitions, from minutes to several hours. Our aim is to characterise lag behaviour over a range of intensity transitions and on timescales particularly relevant to laboratory and industrial XCT, to better understand the limitations of existing models.


The paper proceeds as follows: First we provide the necessary background material including the rate equations from literature modelling detector lag in Sec. \ref{sec:background}. We then describe the experimental method to test and explore these existing rate equations for our application in Sec. \ref{sec:method}. The results of these experiments are presented and analysed in Sec. \ref{sec:results}. The observations as to what detector lag behaviour was unexpected and what was predicted by the models is presented in Sec. \ref{sec:exptObs}. These observations lead us to propose some modifications to the rate equations in Sec. \ref{sec:modifications}. In this section we also provide simulations to show how these changes can capture the unexpected behaviour. The paper then concludes with a discussion of learnings from this research as well as future work in Sec. \ref{sec:discussion} followed by some concluding remarks in Sec. \ref{sec:conclusions}.

\section{Background}
\label{sec:background}

In this section, we review the key theoretical and empirical frameworks used to describe lag in detectors with CsI:Ti scintillators. This background is essential for interpreting our experimental observations and for identifying the limitations of existing models. We first introduce physics-based rate equation models that describe the dynamics of charge carriers and trapping processes within the scintillator. We then present empirical models based on multiexponential impulse responses, which provide a practical description of detector lag without explicitly modelling the underlying physical mechanisms. These approaches form the basis for the analysis and model extensions proposed in this work.

Several theoretical and experimental studies have sought to understand and mitigate afterglow in scintillators, with much of the previous work focusing on scintillation efficiency and detector response on microsecond timescales. Bartram \textit{et al.} developed a theoretical framework based on coupled rate equations to explain the effectiveness of Eu$^{2+}$ doping in suppressing afterglow on millisecond timescales \cite{bartram2006suppression}. Their work showed that deep, long-lived traps can suppress delayed scintillation by scavenging charge carriers that have been released to the conduction band from shallower traps, further delaying radiative recombination pathways. Moretti \textit{et al.} demonstrated good agreement between experimental measurements and a similar, but slightly simplified two-trap rate-equation model for the radioluminescence rise time and afterglow behaviour in YPO$_4$ doped with Ce and Nd \cite{moretti2014radioluminescence}. This framework was later extended to CsI:Tl \cite{moretti2016deep}, where it was shown by experiment and modelling that the presence of deep, long-lived traps can act to suppress afterglow. These studies considered afterglow over timescales of seconds to tens of seconds.

Multiple-trap rate equations from Moretti et al. \cite{moretti2014radioluminescence}, or consistent with those of Bartram et al. \cite{bartram2006suppression} when transients on the microsecond timescale are neglected:
\begin{align}
\frac{d n_c}{dt} &= f_r - \sum_i (N_i - n_i)\,A_{e,i}\,n_c + \sum_i q_i\,n_i - A_r\,n_h\,n_c  \label{eq.moretti1.1}\\
\frac{d n_i}{dt} &= (N_i-n_i)\,A_{e,i}\,n_c - q_i\,n_i, \label{eq.moretti1.2}\\
\frac{d m_v}{dt} &= f_r - (N_h-n_h)\,A_h\, m_v, \label{eq.moretti1.3}\\
\frac{d n_h}{dt}  &= (N_h - n_h)\,A_h\,m_v - A_r\, n_h\, n_c .
\label{eq.moretti1.4}
\end{align}

Here, $N_t$ is the total number of trap types, and $i=1,\ldots,N_t$ labels an individual trap type. The thermal release rate from the $i$-th trap follows the Arrhenius equation:
\begin{align}
q_i = \dfrac{1}{t_i} = s_i \cdot e^{-E_i/kT} 
\label{eq:qi}
\end{align}

\noindent where $n_c$ and $m_v$ are the concentrations of free electrons in the conduction band and
free holes in the valence band, respectively; $n_i$ and $N_i$ are the occupied and total
concentrations of electron traps of type $i$; $n_h$ and $N_h$ are the occupied and total
concentrations of recombination (hole) centres; $A_{e,i}$, $A_h$, and $A_r$ are the
electron-capture, hole-capture, and radiative recombination coefficients, respectively;
$q_i$ is the thermal release probability per unit time from trap type $i$; $t_i$ is the
mean residence time of an electron in trap type $i$; $s_i$ is the frequency (attempt-to-escape)
factor; $E_i$ is the trap depth (activation energy); $k$ is the Boltzmann constant;
$T$ is the absolute temperature; $f$ is the total excitation rate; $\alpha$ is the
fraction of excitation producing prompt scintillation; and
$f_r=(1-\alpha)f$ is the excitation rate contributing to the trapping and recombination
processes.

The emitted scintillation intensity is given by the sum of prompt scintillation and radiative recombination:
\begin{align}
I = \alpha f + A_r \, n_h\, n_c.
\label{eq.moretti-intensity}
\end{align}

Equations (\ref{eq.moretti1.1}) and (\ref{eq.moretti1.2}) describe the time evolution of the electron concentration in the conduction band and in the i-th type of trap. The terms include the generation rate, $f$, of free electron–hole pairs and the direct recombination coefficient, $\alpha$; electron trapping, $(N_i-n_i)\,A_{e,i}\,n_c$, and detrapping, $q_i$, for each trap. The recombination rate is given by $A_r\, m\, n_c$. Equations (\ref{eq.moretti1.3}) and (\ref{eq.moretti1.4}) describe the hole occupancy in the valence band and recombination centre, including hole capture, $(N_h - m)\,A_h\,m_v$.

In parallel with physics-based trapping models, empirical descriptions of detector lag have also been widely used. Starman \textit{et al.} employed a multiexponential representation of the detector impulse response to model lag in amorphous-silicon flat-panel detectors, where the temporal response is expressed as a sum of exponential components associated with different trapping timescales \cite{starman2012nonlinear}:

\begin{align}
    h(t) = b_0 \delta(t) + \sum_{n=1}^{N_{exp}} b_n e^{-a_n t},
    \label{eq.starman-multiexp}
\end{align}
where $t$ is the discrete-time index corresponding to successive x-ray frames, and $N_{exp}$ is the number of exponential terms in the impulse response. $\delta(k)$ is the unit impulse with magnitude 1.0, and the term $b_0\delta(k)$ represents the portion of the input signal unaffected by lag. The coefficients $b_n$ are referred to as the lag coefficients, while the exponential rates $a_n$ are referred to as the lag rates.

Their experiments considered step-up and step-down transitions over a range of exposure intensities, all being either from zero to a finite exposure or from a finite exposure to zero. This approach provides a practical framework for characterising both rising and falling step responses without requiring detailed knowledge of the underlying semiconductor parameters. Their work further demonstrated that lag behaviour depends nonlinearly on exposure intensity which they modelled by allowing the lag rates to vary with exposure level. This formulation provided excellent agreement with experimental observations, although the nonlinear dependence was introduced empirically rather than derived from an underlying physical model. The results also indicated that simple linear time-invariant models may leave residual artefacts in reconstructed images.

\section{Method}
\label{sec:method}


To characterise detector lag over a wide range of transition magnitudes as can be expected to occur in industrial CT and micro-CT scans, the experiments were designed to generate many repeatable step-up and step-down changes in incident x-ray intensity within a single long acquisition. We used a composite step wedge geometry that provides a structured set of attenuation states, while the repeated transitions and embedded clearfield (CF) and darkfield (DF) reference regions allow both lag behaviour and slow source drift to be measured simultaneously. This section describes the experimental design and processing steps used to produce a set of transition curves for subsequent model fitting and comparison.

\subsection{Experimental Setup}
\label{sec:expsetup}
\begin{figure}
\centering

\subfloat[Step wedges of Ti (left) and Cu (right). Thickness of each layer: Ti: 0.1 mm; Cu: 0.08 mm.]{
    \includegraphics[width=.9\columnwidth]{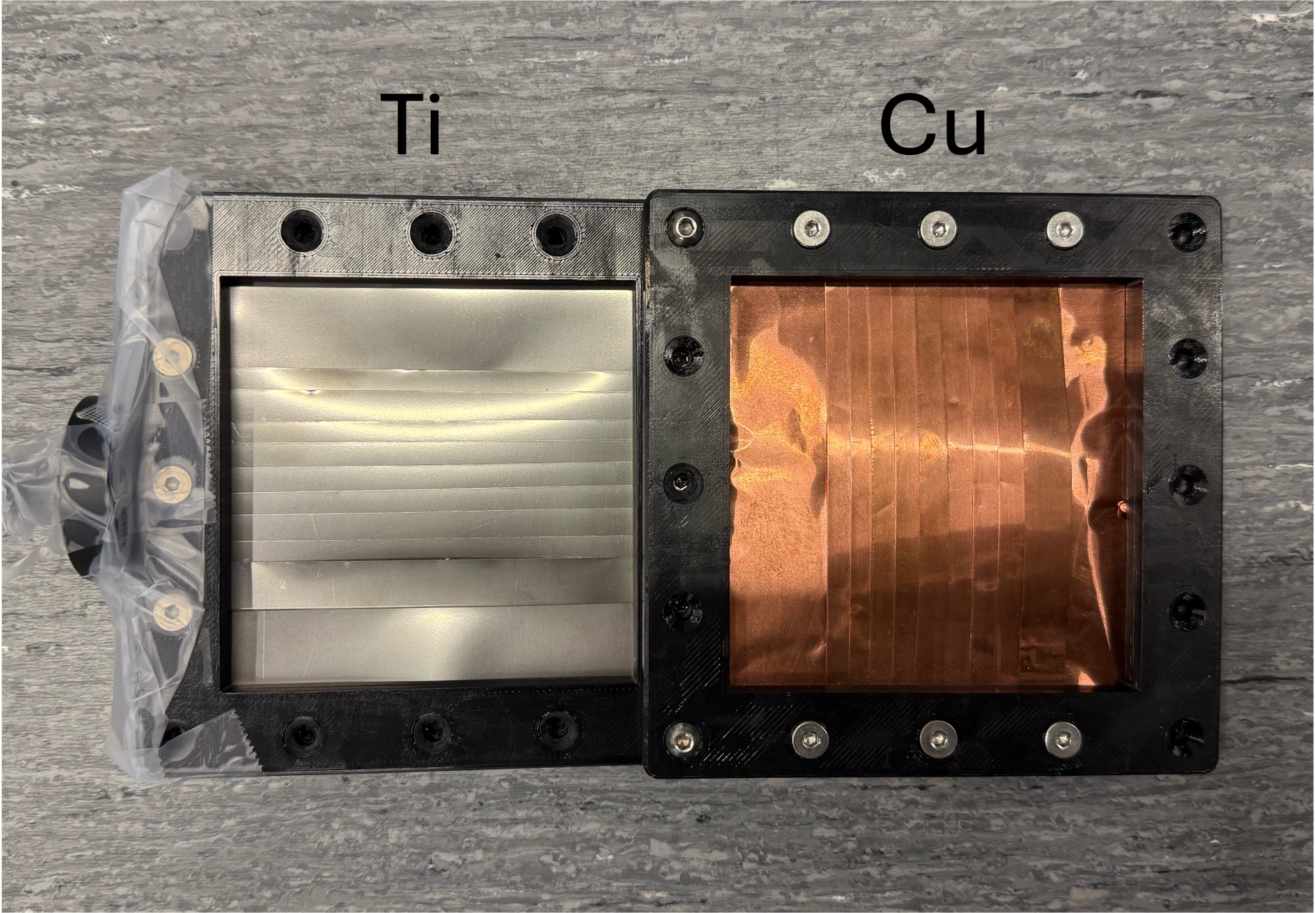}
    \label{fig:fig1a}
}

\vspace{4pt}

\subfloat[Step wedges joined vertically in an orthogonal layout.]{
    \includegraphics[width=.9\columnwidth]{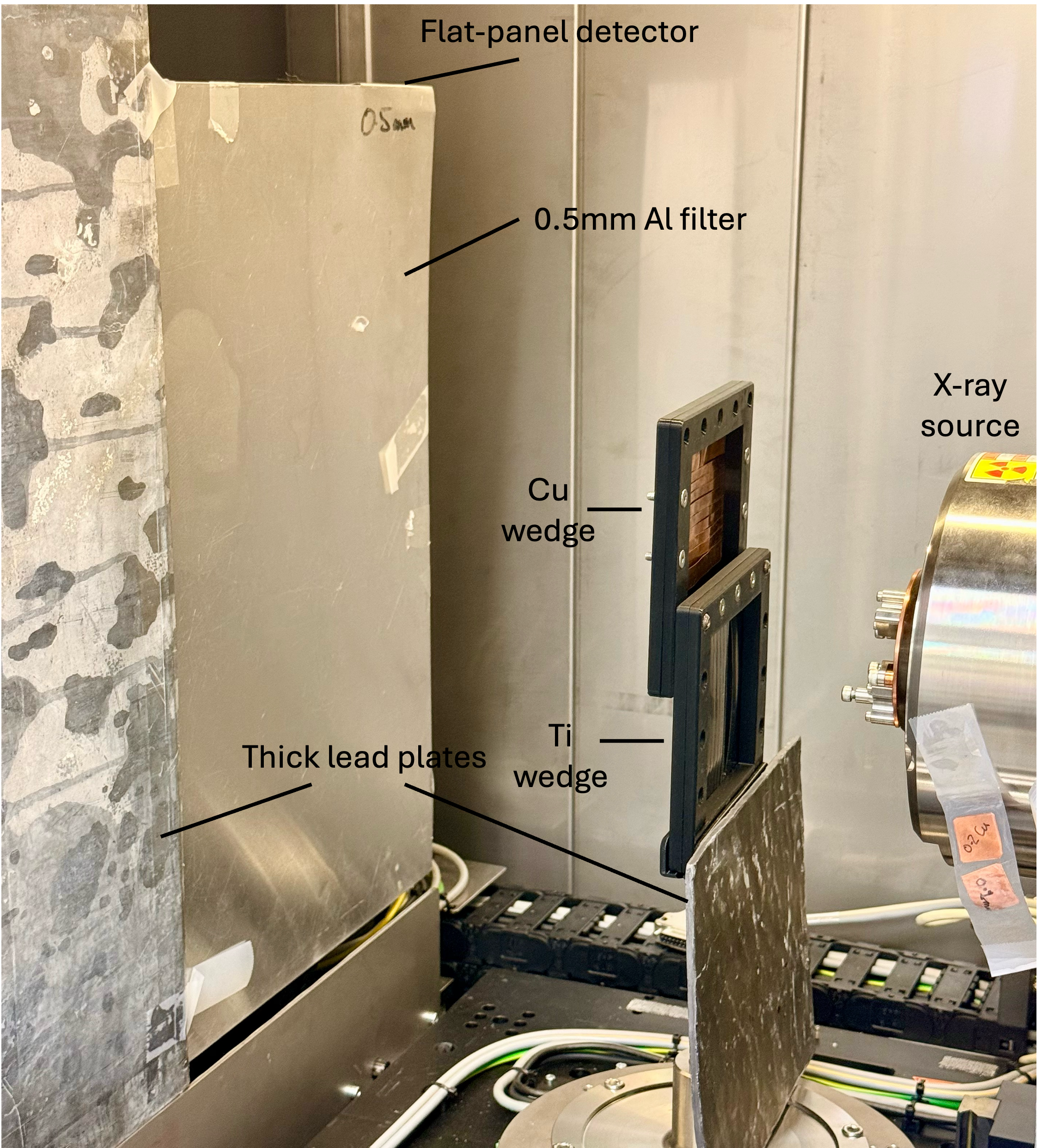}
    \label{fig:fig1b}
}

\caption{Experimental setup of the step wedges for imaging.}
\label{fig:fig1}

\end{figure}

To investigate the behavior of detector lag, including the step-up and step-down relaxation times associated with different intensity ranges, we used two step wedges made of titanium (Ti) and copper (Cu), as shown in Fig. \ref{fig:fig1a}. Both wedges were constructed starting from layer 1 at the bottom, with equal thickness added at each step until reaching the top (layer 10). The thickness of each Ti layer is $(0.10 \pm 0.01)$~mm, while that of each Cu layer is $(0.08 \pm 0.005)$~mm. The two wedges were bound together as shown in Fig. \ref{fig:fig1b}, oriented in orthogonal directions, with a thick piece of lead attached at the very bottom. In addition, the left strip of the detector was covered with a thick lead sheet (> 5 mm).

X-ray images were acquired using the HeliScan MicroCT MkII system (Thermo Fisher Scientific) equipped with a Hamamatsu L10711-03 microfocus x-ray source. The detector distance was set to 350 mm, and the tube voltage was 60 kV, with an exposure time of 2 s. An aperture was placed on the source to reduce secondary radiation from the source, and an approximately 0.5~mm-thick aluminium (Al) filter was placed on the detector.

During image acquisition, the x-ray source remained on continuously since we found that x-ray ouput was less stable for some period of time after being turned on. Thick lead plates ($\sim 4~\mathrm{mm}$) were used to mimic the dark field (DF) condition in this experiment. The acquisition consisted of seven sequential stages, corresponding to different material configurations: CF0 (clear field), Cu0 (moving to the Cu wedge), Ti0 (moving to the Ti wedge), Cu1, Ti1, Cu2 (repeating once and then returning to the Cu wedge), and DF1 (finally returning to the DF condition to observe afterglow), with 4000 images recorded in each stage. Each image was obtained by averaging five consecutive frames, resulting in an effective exposure time of 12 s per image and a total acquisition time of approximately 93 hours, with each stage lasting more than 12 hours. 

\begin{figure*}[!htbp]
\centering
\includegraphics[width=\textwidth]{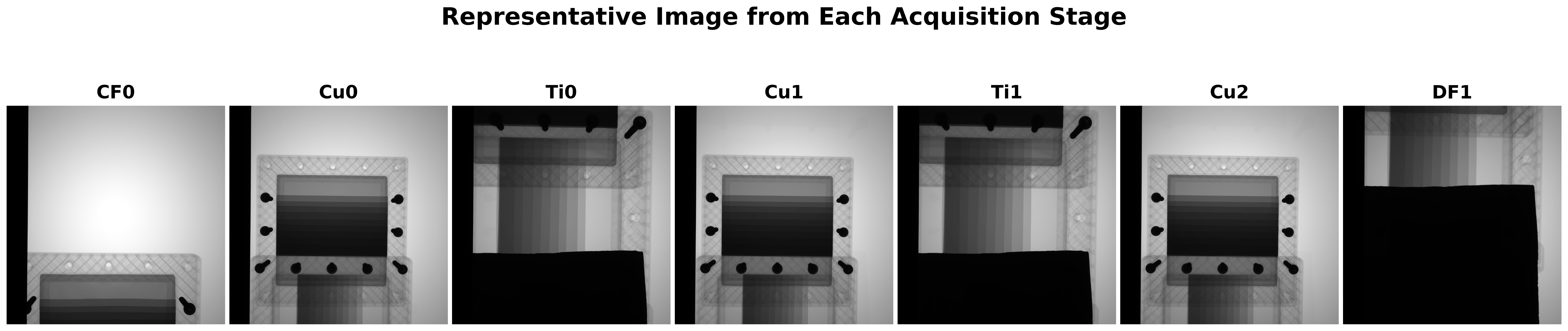}
\caption{Representative corrected X-ray images from the seven acquisition stages: clear field (CF0), Cu0, Ti0, Cu1, Ti1, Cu2, and dark field (DF1). Each panel shows the same field of view under different material configurations.}
\label{fig:correctedimgs}
\end{figure*}

One representative image from each stage is shown in Fig. \ref{fig:correctedimgs}. This setup enabled us to obtain a $10 \times 10$ grid of transitions across different intensity levels. For example, the top-left square corresponds to a sequence of transitions from the CF region to the thinnest Cu step, followed by the thickest Ti step, with this cycle repeated once before finally transitioning to the DF region.

\begin{figure}
\centering
\includegraphics[width=.7\linewidth]{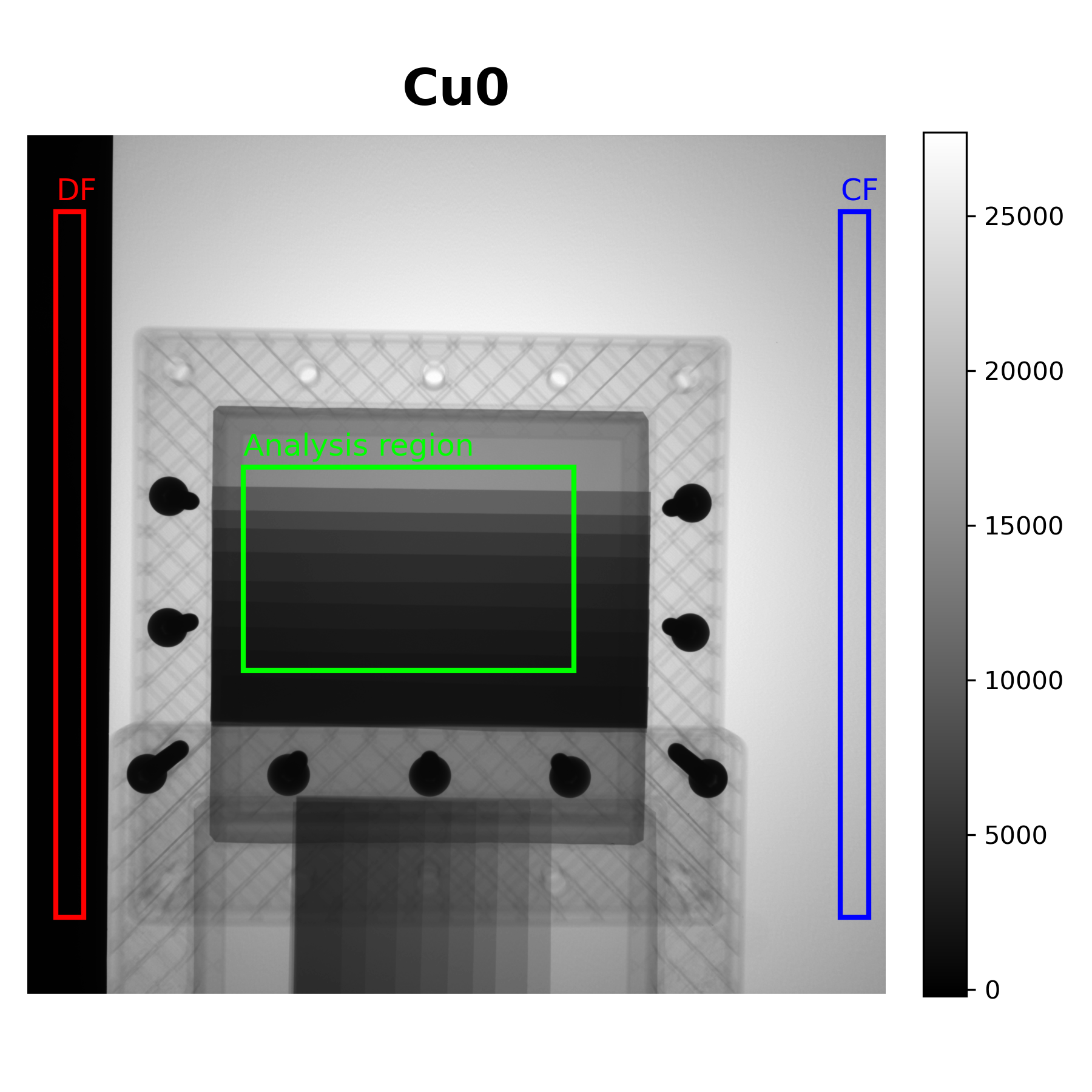}
\caption{Example corrected image from the first Cu stage. The red box indicates the dark field (DF) reference region, the blue box indicates the clear field (CF) reference region, and the green box shows the analysis region used for extracting all transition curves curves throughout the experiment.}
\label{fig:Cu0}
\end{figure}

We intentionally left a CF stripe on the right side of the detector and a DF stripe on the left throughout the acquisition process (see Fig.~\ref{fig:correctedimgs}). During such long acquisitions, these reference regions were used to correct for source instabilities, such as sudden intensity jumps and other fluctuations during imaging. Figure~\ref{fig:Cu0} shows an example image from the first Cu stage, where the analysis region used throughout the experiment is indicated by the green box. This region was corrected using

\begin{equation}
I_{\text{corrected}} = (I - \mathrm{DF}) 
\times 
\frac{\mathrm{CF}_{\text{avg}} - \mathrm{DF}_{\text{avg}}}
{\mathrm{CF} - \mathrm{DF}},
\label{eq:cf_df_correction}
\end{equation}

\noindent where $I$ is the raw intensity, $\mathrm{DF}$ is the average intensity from the red boxed region, $\mathrm{CF}$ is the average intensity from the blue boxed region, $\mathrm{DF}_{\text{avg}}$ is the average DF over all 28,000 images, and $\mathrm{CF}_{\text{avg}}$ is the average CF over all 28,000 images.

\subsection{Data Processing}
\label{sec:fitmethod}

As mentioned in Sec.~\ref{sec:expsetup}, we obtained a $10 \times 10$ grid of transition data. Due to the restricted analysis region (green box in Fig.~\ref{fig:Cu0}), the thickest (10th) layer of the Cu wedge was excluded, resulting in an effective $9 \times 10$ grid. Within each grid cell, we extracted the mean corrected intensity of the central $30 \times 30$ pixel region for analysis.

\begin{figure*}
    \centering
    \includegraphics[width=.9\linewidth]{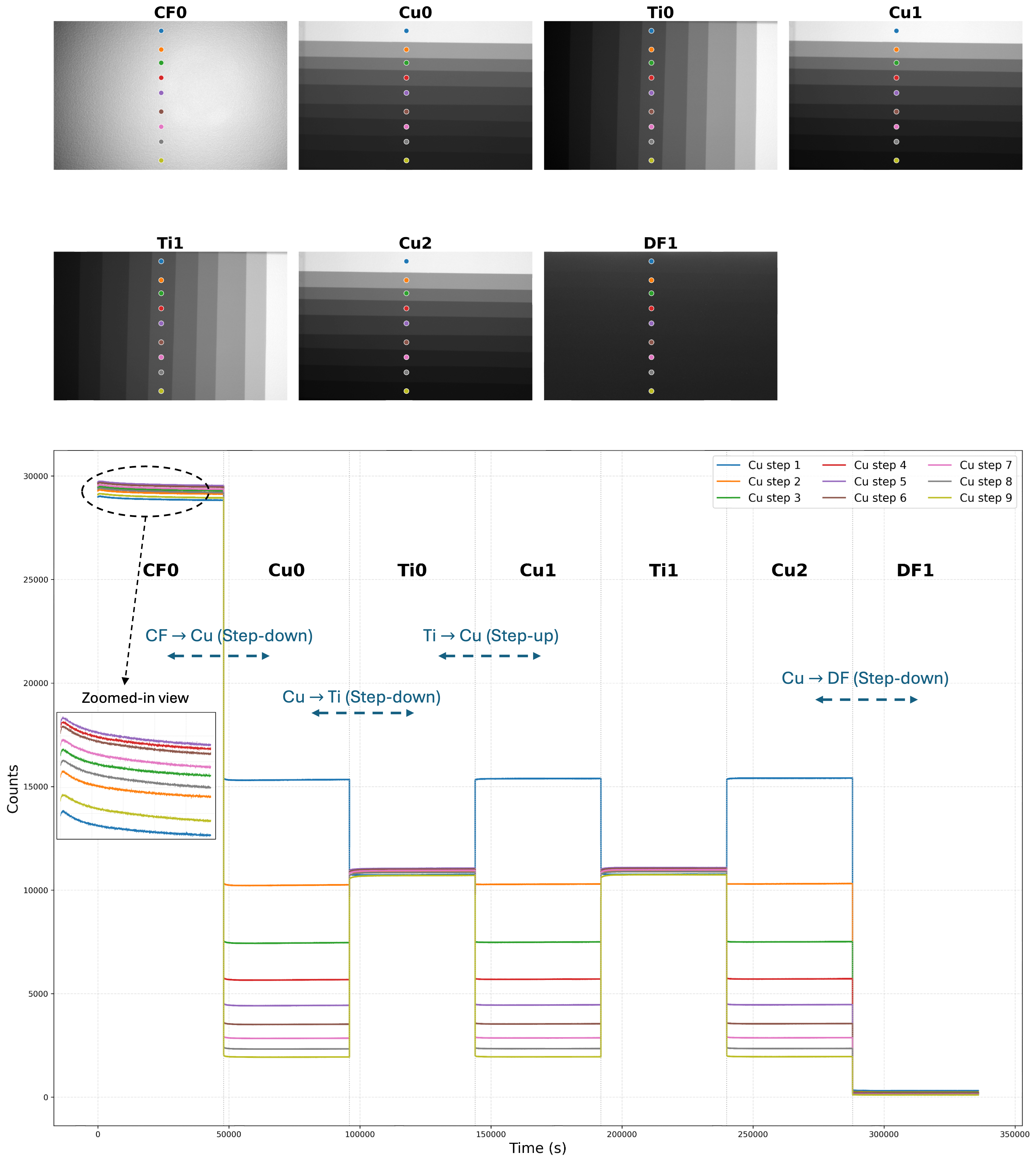}
    \caption{Representative intensity profiles over all seven acquisition stages (CF0, Cu0, Ti0, Cu1, Ti1, Cu2, and DF1). The mean intensity curves are extracted at a fixed horizontal position ($x$), corresponding to the same Ti column, while varying the Cu step positions. Vertical dashed lines indicate stage boundaries, and the four main transition types are highlighted: CF$\rightarrow$Cu (step-down), Cu$\rightarrow$Ti (step-up), Ti$\rightarrow$Cu (step-down), and Cu$\rightarrow$DF (step-down). The seven images shown above the profile indicate the acquisition geometry at each stage. Each stage is zoomed into the central analysis region, as marked in Fig. \ref{fig:Cu0}. The coloured dots mark the exact pixel locations from which the profiles are extracted, demonstrating that all curves originate from the same Ti layer, while sampling different Cu step heights. A zoomed-in inset of the CF0 image is included for clarity, showing the shape of the initial CF region. Here, ``counts'' refer to the detector output in analogue-to-digital units (ADUs).}
    \label{fig:expprofile}
\end{figure*}

Figure~\ref{fig:expprofile} shows representative intensity profiles extracted from $30\times30$ pixel ROIs over the full seven-stage acquisition using a fixed column at $x=1300$ pixels. Each curve corresponds to a different Cu step position (nine levels) while sampling the same Ti layer. The corresponding sampling locations are marked by colored dots in the seven stage images (CF0, Cu0, Ti0, Cu1, Ti1, Cu2, DF1), highlighting that all profiles are extracted from the same Ti column across the acquisition sequence. Variations in the absolute intensity levels across curves, even within the same Ti material step, are attributed to changes in the cone-beam source-to-detector distance across the field of view. Stage boundaries and the main transition types (CF$\rightarrow$Cu, Cu$\rightarrow$Ti, Ti$\rightarrow$Cu, and Cu$\rightarrow$DF) are annotated. Since Cu is generally more attenuating than Ti under our experimental conditions, for example, two Cu sheets produce greater attenuation than eight Ti sheets, we mainly focused on the following four transition stages, as also indicated in Fig.~\ref{fig:expprofile}:

\begin{enumerate}
    \item CF $\rightarrow$ Cu: intensity step-down, transitioning from clear field to Cu.
    \item Cu $\rightarrow$ Ti: intensity step-up, transitioning from lower to higher intensity for almost all squares.
    \item Ti $\rightarrow$ Cu: intensity step-down, transitioning from higher to lower intensity for almost all squares.
    \item Cu $\rightarrow$ DF: intensity step-down from active exposure toward the dark field (afterglow).
\end{enumerate}

One repetition of transitions (2) and (3) was included to assess the repeatability of the detector response and to investigate the presence of long-lived charge traps, which would be expected to progressively fill over many hours and therefore to be in a different state between successive repetitions of the same transition. Note that although the majority of the behaviour within each transition stage exhibits the same step-up/step-down trend, a small number of cases show the opposite (e.g., several step-down curves in transition (2): intensity step-up). In the analysis, we omit these cases for clarity, as they exhibit the same overall step-up/step-down behaviour as the main transition stages we focus on. 

For each transition stage, we plotted the intensity decay or growth over time using the following expression, with the curves colored by the absolute intensity difference $|I_0 - I_f|$:

\begin{equation}
I(t) - I_f ,
\label{eq:unnorm}
\end{equation}

where $I(t)$ is the measured intensity at time $t$, $I_0$ is the initial intensity, and $I_f$ is the final intensity. For example, when analysing the transition from Ti to Cu, $I_0$ is defined as the average intensity over the last 6 minutes of the Ti segment, and $I_f$ is the corresponding average over the Cu segment.

\section{Experimental results}
\label{sec:results}
\begin{figure*}
    \centering
    \includegraphics[width=.9\linewidth]{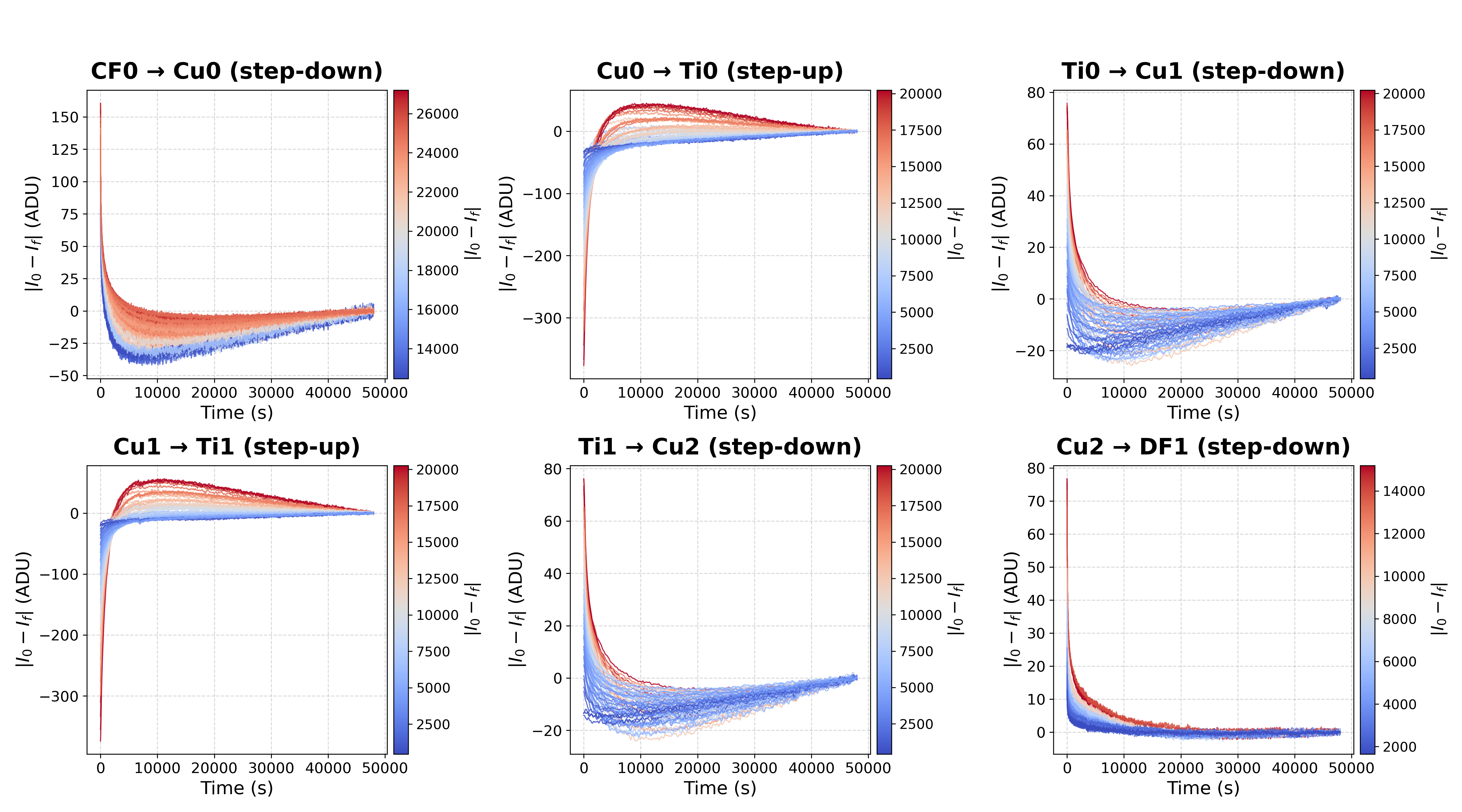}
    \caption{Intensity transition curves, $|I(t) - I_f|$ in ADU, for the four main transition types (CF$\rightarrow$Cu, Cu$\rightarrow$Ti, Ti$\rightarrow$Cu, and Cu$\rightarrow$DF), colored by the absolute transition intensity difference $|I_0 - I_f|$. 100~ADU is $\sim0.3\%$ of the CF intensity and $\sim5\%$ of the largest step wedge intensity (Cu step~9).}
    \label{fig:intensityplots}
\end{figure*}

Figure~\ref{fig:intensityplots} shows the intensity transition curves, coloured by the absolute transition intensity difference $|I_0 - I_f|$, with red indicating larger intensity changes and blue indicating smaller intensity changes. Several key features are observed:

\begin{itemize}
    \item Overshoot on a timescale of several hours is observed in the Cu$\rightarrow$Ti (step-up) transitions, where the transient response exceeds the final value $I_f$. The overshoot increases with increasing $|I_0-I_f|$, reaching approximately 50--60 analog-to-digital units (ADU) for the largest transitions (red curves).

    \item The temporal shape of the step-up overshoot closely resembles that observed in the CF0 intensity profile (Fig.~\ref{fig:expprofile}), although the CF0 profile is shown only for qualitative comparison since its initial state is not well controlled in this experiment.

    \item Overshoot is also observed in the Ti$\rightarrow$Cu (step-down) transitions, where the transient response falls below the final value $I_f$. Unlike step-up overshoot, the step-down overshoot increases with decreasing $|I_0-I_f|$, reaching approximately 20--25~ADU for the smallest transitions (blue curves).

    \item A consistent rising trend is observed in the low-intensity transition curves (referred to hereafter as the ``rising baseline'', shown in blue in these plots), corresponding to transitions with relatively small intensity changes. The rising baseline may be sufficient to capture the observed step-down overshoot and may represent the same underlying phenomenon.
    
    \item The slope of this rising baseline decreases progressively over the duration of the experiment.
    
    \item The final transition to DF shows a complete decay of the signal with relatively little afterglow on the timescale of several hours.
\end{itemize}

To further quantify the observed behaviour, we fitted the multi-exponential model (Eq.~\ref{eq.starman-multiexp}) to all transition curves obtained from intensity changes (Cu$\rightarrow$Ti and Ti$\rightarrow$Cu). Four exponential terms were used, with the lag rates ($a_n$) determined via a global fit, followed by individual fits of the lag coefficients $b_n$ for each curve, with all $b_n$ allowed to take negative values to capture overshoot behaviour. 

This is in contrast to the model proposed in \cite{starman2012nonlinear}, where the lag rates were allowed to vary with transition intensity and only positive values of $b_n$ were permitted. With these modifications, the model provided excellent agreement with the experimental data, achieving a mean coefficient of determination ($R^2$), averaged across all fitted curves, of 0.985 for the step-up transitions and 0.957 for the step-down transitions.

We also explored extending the model to five exponential components. However, this did not lead to a noticeable improvement in fitting quality. Instead, it introduced increased variability in the fitted coefficients, with less consistent trends across different transitions. 

The globally fitted time constants (i.e., $\tau_n = 1/a_n$) are $\tau_1 = 125$ s, $\tau_2 = 1066$ s, $\tau_3 = 4279$ s, and $\tau_4 = 100000$ s, where the last value reaches the fitting algorithm limit.

\begin{figure*}
    \centering

    \begin{subfigure}{0.495\linewidth}
        \centering
        \includegraphics[width=\linewidth]{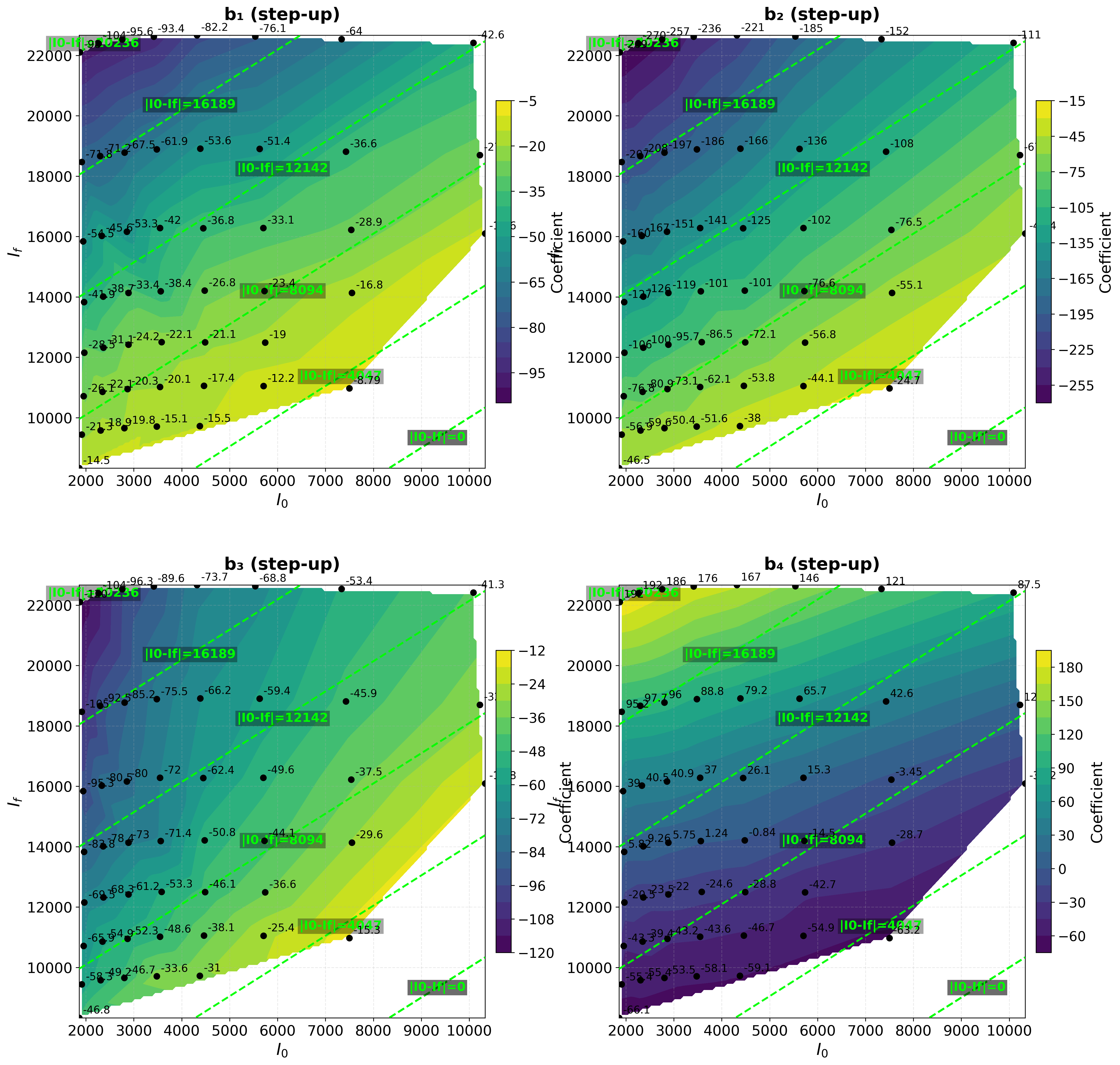}
        \caption{Step-up coefficients $b_1$--$b_4$.}
        \label{fig:step-up_contours}
    \end{subfigure}
    \hfill
    \begin{subfigure}{0.495\linewidth}
        \centering
        \includegraphics[width=\linewidth]{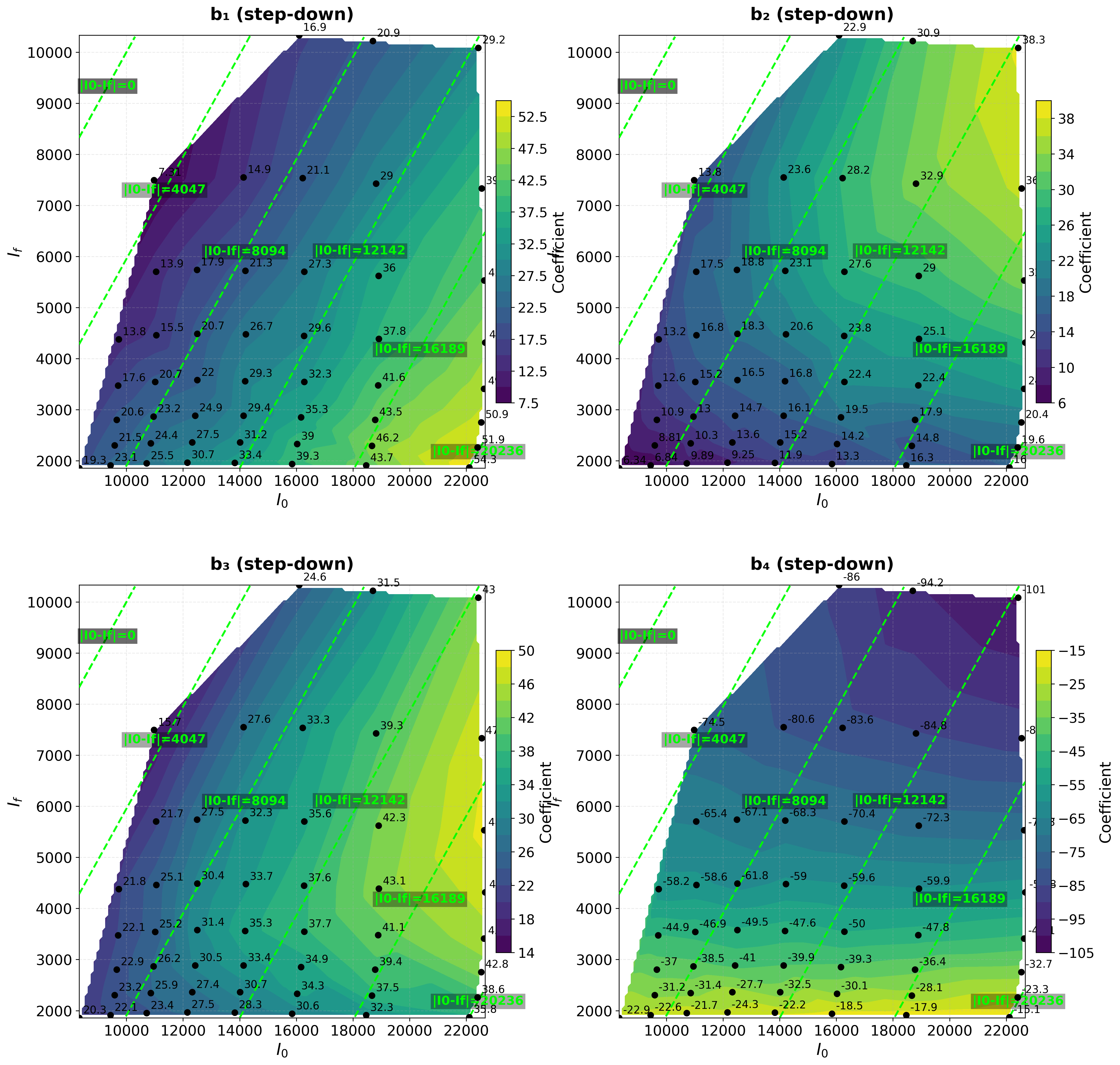}
        \caption{Step-down coefficients $b_1$--$b_4$.}
        \label{fig:step-down_contours}
    \end{subfigure}

    \caption{Contour plots of lag coefficients for step-up (left) and step-down (right) transitions as a function of $(I_0, I_f)$. Iso-lines of $|I_0 - I_f|$ are overlaid in each plot.}
    \label{fig:bn_contours}
\end{figure*}

Figure~\ref{fig:bn_contours} shows the contour plots of the coefficients $b_n$ for both step-up and step-down transitions as a function of the initial and final intensities ($I_0$, $I_f$). Iso-lines of $|I_0 - I_f|$ are overlaid on each plot to highlight the dependence on transition magnitude.

\begin{itemize}
    \item The first three coefficients ($b_1$--$b_3$) are approximately functions of the transition magnitude $|I_0-I_f|$ alone, exhibiting similar trends for both step-up and step-down transitions.
    
    \item The coefficient associated with the longest time constant ($b_4$) exhibits a different trend compared to the other coefficients:
    \begin{itemize}
        \item In the step-down case, $b_4$ has an opposite sign to $b_1$--$b_3$.
    
        \item In the step-up case, $b_4$ changes sign from negative to positive.
    \end{itemize}
    
\end{itemize}

\section{Observations and interpretation from experimental results}
\label{sec:exptObs}


The observed overshoot behaviour indicates that the detector response cannot be described by a simple linear relaxation process, and is inconsistent with existing rate equation models. 

The rising baseline observed across all datasets and the step-down overshoot are consistent with gradual charge accumulation in a deep radiative trap, leading to incomplete relaxation between exposures. The decreasing slope over time suggests a reduction in the effective filling rate, which may arise from equilibration as a radiative trap recombination rate approaches a constant value, consistent with existing models, or from nonlinear recombination kinetics of a radiative or non-radiative trap.

The presence of overshoot in the largest-magnitude step-up transitions is particularly notable. This behaviour is not readily explained by rate equation models of purely radiative traps, and will be further discussed in later sections.

The relatively short-lived afterglow observed in the final transition to DF suggests that long-lived radiative traps are not dominant. Instead, this behaviour is more consistent with non-radiative deep traps that efficiently remove carriers from shallow traps.

The unique behaviour of $b_4$, compared with $b_1$--$b_3$, indicates that the longest time constant is associated with more than one physical mechanism. While $b_1$--$b_3$ exhibit relatively smooth and similar trends across both step-up and step-down transitions, $b_4$ shows a much stronger dependence on the transition direction and the exposure levels. In particular, $b_4$ changes sign between different step-up transitions and has the opposite sign to $b_1$--$b_3$ during step-down transitions. 

In the step-down case, $b_4$ has a different sign compared to the first three coefficients, indicating that it governs the step-down overshoot (``rising baseline'') behaviour. It is also approximately proportional to $I_f$, which is consistent with the effect of a steadily filling deep trap.

In the step-up case, $b_4$ changes sign from negative to positive, corresponding to the transition from a rising (blue) baseline to a (red) overshoot. The rising baseline is consistent with a steadily filling deep trap, as in the step-down case, whereas the overshoot for larger steps ($|I_0-I_f|$) is more complex to interpret.

This suggests that the longest-timescale component cannot be explained solely by delayed relaxation towards equilibrium, but instead reflects multiple physical processes whose relative contributions depend on the transition conditions. The origin of this behaviour, particularly the overshoot observed in the largest step-up transitions without an equivalent response during step-down transitions, is not readily explained by the present model and will be discussed in later sections.

\section{Proposed modifications to rate equations}
\label{sec:modifications}


The multiple-trap rate equation model proposed in \cite{moretti2014radioluminescence}, and similarly in \cite{bartram2006suppression}, provides a physically motivated description of detector lag based on carrier trapping and thermal release. While these models successfully describe the long-term decay associated with trapped charge populations, they cannot readily explain the pronounced overshoot observed during large step-up transitions, which implies a transient depletion process. These observations suggest that additional physical processes beyond trapping, detrapping, and radiative recombination contribute to the detector response. 

\subsection{Model}
\label{sec:6.1model}

To account for the observed behaviour, we propose a modified version of the multiple-trap rate equation model, with several additional terms and modifications indicated in red and green.

The proposed multiple-trap rate equations are:
\begin{align}
\frac{d n_c}{dt} &= f_r - \sum_i (N_i - n_i)\,A_{e,i}\,n_c + \sum_i q_i\,n_i \notag\\
&\quad - A_r\,{n_h}^{\color{darkred}\beta}\,n_c - {\color{darkgreen} A_{nr}\,n_h\,n_c\,n_5},
\label{eq.proposed1.1}\\[4pt]
\frac{d n_i}{dt} &= (N_i-n_i)\,A_{e,i}\,n_c - q_i\,n_i - {\color{darkred} A_{nr,i}\,n_h\,n_i},
\label{eq.proposed1.2}\\[4pt]
\frac{d m_v}{dt} &= f_r - (N_h-n_h)\,A_h\,m_v,
\label{eq.proposed1.3}\\[4pt]
\frac{d n_h}{dt} &= (N_h - n_h)\,A_h\,m_v - A_r\,{n_h}^{\color{darkred}\beta}\,n_c \notag\\
&\quad - {\color{darkred} \sum_i A_{nr,i}\,n_h\,n_i} - {\color{darkgreen} A_{nr}\,n_h\,n_c\,n_5}.
\label{eq.proposed1.4}
\end{align}

\noindent with the emitted scintillation intensity combining prompt emission and radiative recombination:
\begin{align}
    I = \alpha f + A_r \, {n_h}^{\color{darkred}\beta} \, n_c.
    \label{eq.proposedI}
\end{align}

All existing terms are as previously defined in Sec.~\ref{sec:background}. These are slightly generalised from that in \cite{moretti2014radioluminescence} (also similar to that in \cite{bartram2006suppression}), by the introduction of the following to accommodate the observations:
\begin{enumerate}
    \item An exponent $\beta$ for $m$ in the recombination term, indicative of fractional-order kinetics, set to $\beta = 0.5$. This value was consistently found by the optimiser. It is further supported by the observed scaling of the trap emission intensity (and therefore the trap populations), which follows approximately $I \propto f^{2/3}$, consistent with the results in \cite{starman2012nonlinear}, who found that the trap occupancy scales as $N_i \propto f^{2/3}$, again implying a delayed scintillation intensity of $I \propto f^{2/3}$. See Appendices~\ref{sec:appendix_steady_state} and \ref{sec:appendix_beta_evidence} for the derivation showing why $\beta = 0.5$ gives $I \propto f^{2/3}$.
    \item A non-radiative recombination term is introduced for trap~4 ($i=4$), controlled by $A_{nr,4}$, and is implemented through the red terms in Eqs.~\ref{eq.proposed1.2} and \ref{eq.proposed1.4}. We choose the deepest traps,~4 and~5, to have non-zero $A_{nr,i}$, and to be fully non-radiative by setting $q_i$ to zero. It is introduced because a long-lived radiative trap is inconsistent with the observations of near-linear growth during exposure (following both step-up and step-down) on the $\sim$12~hour timescale, but no corresponding afterglow decay mode on this timescale in the Cu2$\rightarrow$DF1 curves. 
    \item A fifth trap is introduced that nonlinearly reduces the overall scintillation efficiency through radiationless recombination between conduction electrons and holes, determined by a nonlinear depletion term indicated in the green terms in Eqs.~\ref{eq.proposed1.1} and \ref{eq.proposed1.4}. The rate of this radiationless recombination is proportional to the number of conduction electrons, $n_c$, the number of holes, $m$, the number of electrons in trap~5, $n_5$, and the depletion coefficient, $A_{nr}$. This additional term can account for the nonlinear ``overshoot'' observed during large step-up transitions on the $\sim$12~hour timescale. The absence of a corresponding overshoot during step-down transitions suggests that this process depends primarily on the final exposure level rather than the transition magnitude $|I_0-I_f|$. This assumption is consistent with the experimental data, however, we are not aware of a specific physical motivation for this nonlinear depletion trap.
\end{enumerate}

\subsection{Simulation Method}
\label{sec:modSimulationmethod}


The modified rate-equation model was implemented numerically by solving Eqs.~\ref{eq.proposed1.1}--\ref{eq.proposed1.4}. The experimentally measured sequence of exposure intensity conditions (CF0 $\rightarrow$ Cu0 $\rightarrow$ Ti0 $\rightarrow$ Cu1 $\rightarrow$ Ti1 $\rightarrow$ Cu2 $\rightarrow$ DF1) was used directly as the driving function, $f$, and the simulated scintillation intensity (Eq.~\ref{eq.proposedI}) was calculated throughout the entire 7-stage exposure history.

The characteristic lifetimes of the first four traps were taken from Section~\ref{sec:results}, where they were estimated by fitting multiple exponential components to the experimental data and subsequently used as the midpoints of the parameter bounds (limits) for bounded optimisation using the DIRECT global optimiser~\cite{jones1993lipschitzian} (see Appendix~\ref{sec:appendix_model_optimisation} for more details). As described in Sec.~\ref{sec:6.1model}, to account for the additional long-timescale features observed in the measurements, a deep non-radiative trap (trap 4) was introduced to model the rising baseline, while a fifth trap was introduced to model the overshoot observed during the step-up transitions.

The parameter search was performed iteratively by minimising the discrepancy between the simulated and experimental transition curves over all six step-up and step-down stages simultaneously. The fitting procedure was guided primarily by reproducing the observed amplitudes, the step-up and decay timescales, and the nonlinear overshoot behaviour.\

\subsection{Simulation Results and Discussion}
\label{sec:modSimulationresults}

The optimised parameter values are broadly consistent with those reported in \cite{moretti2016deep}, the complete set of optimised parameters is provided in Appendix~\ref{sec:appendix_model_optimisation}. Figure~\ref{fig:exp_vs_sim} compares representative experimental and simulated curves selected across the full range of $|I_0-I_f|$. Across all transitions, the optimised model captures about 90\% of the lag, achieving a root mean square error of 9~ADU in $|I_0-I_f|$. This corresponds to an accuracy of approximately 1~ADU in the relaxation curves, or around 0.019\% of the CF exposure level. Overall, the modified model reproduces the main qualitative features of the measurements. In particular:

\begin{enumerate}
    \item The model captures the rapid initial response followed by the slow recovery observed in most transitions.
    \item The step-up transitions exhibit a pronounced overshoot in both experiment and simulation for large $|I_0-I_f|$ (red curves), supporting the inclusion of the nonlinear depletion term involving trap 5 in Eqs.~\ref{eq.proposed1.1} and \ref{eq.proposed1.4}.
    \item The blue curves, corresponding to the smallest intensity changes, exhibit the same rising baseline behaviour in the step-down transitions in both experiment and simulation, supporting the inclusion of trap 4.
\end{enumerate}

\begin{figure*}[t]
    \centering
    \includegraphics[width=\textwidth]{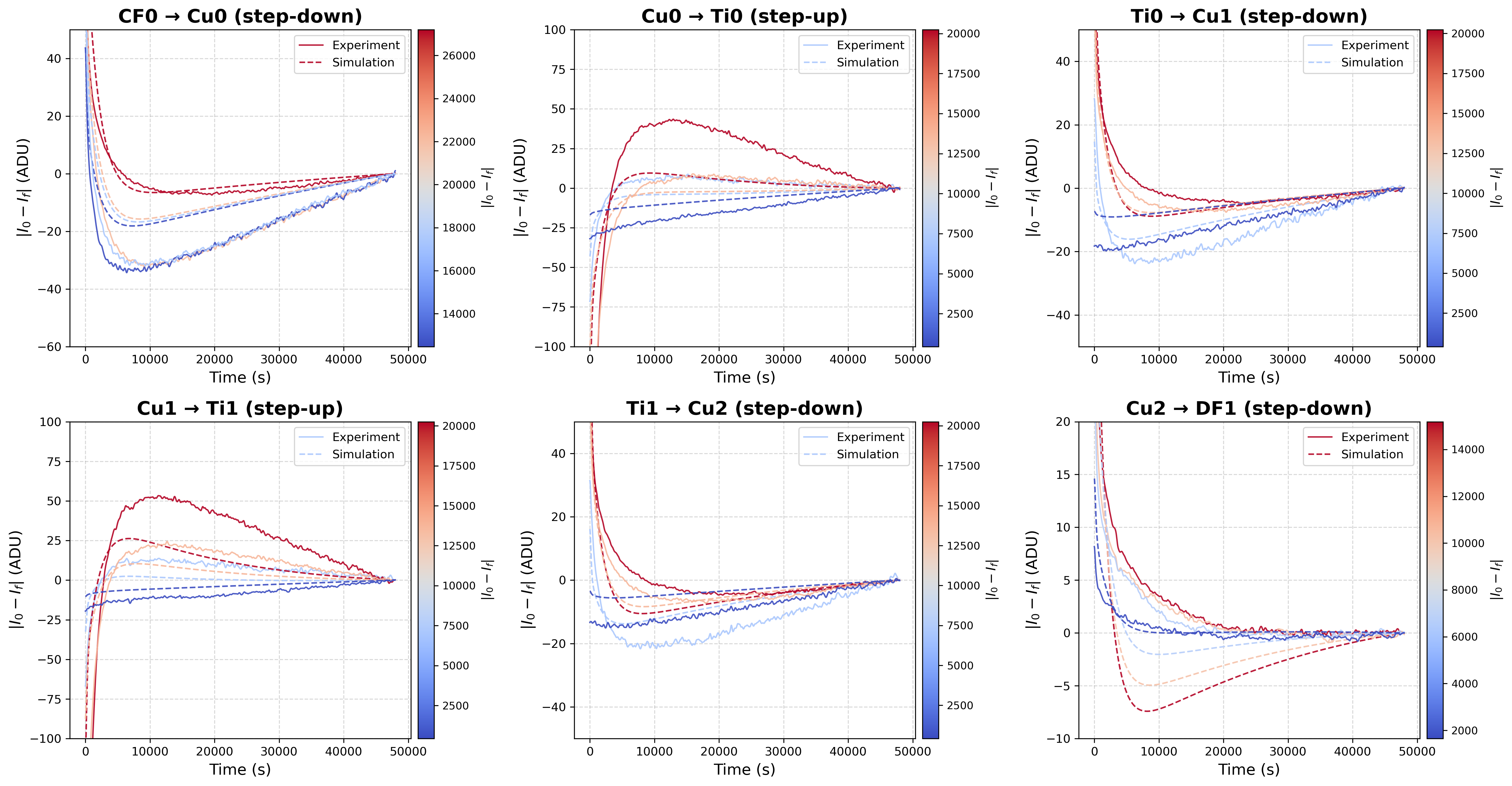}
    \caption{Comparison between experimental (solid lines) and simulated (dashed lines) transition curves for the six transition stages. Curves of different colours correspond to transitions spanning the full measured range of $|I_0-I_f|$, with the colour scale indicating the magnitude of the intensity change. The overall RMSE between the experimental and simulated curves is 9~ADU.}
    \label{fig:exp_vs_sim}
\end{figure*}

However, some discrepancies remain: the simulations generally underestimate the magnitude of the overshoot during the step-up transitions and do not perfectly reproduce the long-time tails of several curves; for the afterglow transition ($\mathrm{Cu2}\rightarrow\mathrm{DF1}$), the experimental data exhibit no reverse overshoot behaviour, whereas the simulation still predicts a rising baseline, as it does for normal step-down transitions.

Therefore, our results support the conclusions in \cite{bartram2006suppression} and \cite{moretti2014radioluminescence}, that afterglow is suppressed by a deep trap. However, there are two additional learnings:

\begin{itemize}
\item Deep trap~4 needs to be non-radiative, otherwise there will be a ``falling baseline'' in the afterglow curves with a similar slope to the rising baseline. This is not consistent with the observations, in which there is no multi-hour afterglow.

\item The deep trap does not significantly suppress radiation from the shallower traps in the step-up and step-down transitions, the suppression only occurs for afterglow. This is explained by the model: the suppression is captured in the rate equations by competition between the scavenging term, $A_{e4}(N_4-n_4)n_c$, and the radiative recombination term, $A_r m^\beta n_c$. The radiative term scales with $m^\beta$, with $\beta=1$ in previous models and $\beta\approx0.5$ being optimum in our model. Since typical values of $m$ (the number of holes) during exposure are $\sim10^{12}$, radiative recombination will tend to dominate during exposure, while the scavenging term may dominate during afterglow. With the model values of $A_r\approx10^{-6}$ and $A_{e4}N_4\approx10^2$, it is clear that the scavenging term will dominate for $m\lesssim10^8$. Therefore, radiative recombination will tend to dominate for significant exposure levels, whereas at much lower exposure, corresponding to the afterglow case, the scavenging term may dominate, suppressing afterglow. Therefore, this suppression does not occur during intensity transitions that will be seen during tomographic imaging.
\end{itemize}

\section{Discussion and future work}
\label{sec:discussion}
This work presents a robust and repeatable experimental framework for characterising lag in scintillation-type x-ray detectors over long time scales by radiographic imaging of transitions between two step wedges. By incorporating both clear-field (CF) and dark-field (DF) regions in each radiograph, the measured signals were corrected for x-ray intensity fluctuations. The resulting transition curves exhibit features that have not been observed previously, such as large overshoots that were highly repeatable across a range of measurements. These observations do not conflict with previous studies, but rather arise because the present work investigates a wider range of transition magnitudes and much longer timescales than have previously been examined. The high repeatability of these features strongly suggests that they originate from the detector response, providing confidence in the quality and reproducibility of the experimental methodology.

Despite the CF--DF correction, as described in Sec.~\ref{sec:expsetup}, a small systematic bump remained in two of the measured step-up transitions. In the present work, this feature was compensated for by manually shifting the affected data to align the transitions, and this correction is not expected to affect the overall trends or the subsequent analysis. The exact origin of this behaviour is currently unknown and may arise from detector synchronisation, x-ray source fluctuations, calibration errors, or other experimental factors. A more detailed investigation of this effect should be conducted in future work to eliminate the need for manual correction and further improve the robustness of the measurement procedure.

The measured transition curves were accurately represented using a multi-exponential fitting approach similar to that used in \cite{starman2012nonlinear}. The fitted curves closely matched the experimental data, and the fitted lag coefficients exhibited smooth, consistent trends across different transitions (see Fig.~\ref{fig:bn_contours}), further demonstrating the high quality of the measurements. Furthermore, the fitted global time constants provide reliable estimates of the detector lag rates, which were subsequently used as inputs to the proposed multiple-trap rate equation lag model.

To better reproduce the observed detector behaviour, several modifications were introduced to existing multiple-trap rate equation lag models \cite{bartram2006suppression,moretti2014radioluminescence}, including a non-radiative recombination term for the traps and a nonlinear depletion term, as described in Section~\ref{sec:6.1model}. Estimating the model parameters proved challenging because of the large parameter space, differences in parameter scales, and the presence of multiple local minima (with the process described in Appendix~\ref{sec:appendix_model_optimisation}). Consequently, parameter estimation relied on exponential fitting to set the parameter bounds, followed by bounded optimisation. A wide range of parameter combinations produced similarly good overall fit metrics, indicating that the optimisation converged to a compromise between competing features of the experimental data rather than a unique physical solution. The modified model successfully reproduced the overall temporal trends of the measured transitions, particularly the overshoot behaviour that initial models \cite{bartram2006suppression,moretti2014radioluminescence} did not account for. However, discrepancies remain. The simulations underestimate the magnitude and duration of the overshoot during step-up transitions and fail to reproduce certain behaviours, such as the afterglow transition. These observations suggest that the model is imperfect and that a new model for overshoot, in particular, is needed.

Overall, this work provides a significantly improved understanding of detector lag behaviour for transition intensities and timescales relevant to industrial and micro-CT, and establishes a reliable experimental methodology for its characterisation. Although the proposed mathematical model is not yet sufficiently complete for direct lag correction in practical imaging systems, the experimental insights obtained here can already be used to design acquisition protocols that minimise lag-related artefacts. Future work will focus on refining the mathematical model by considering additional physical mechanisms that can better reproduce the observed overshoot and afterglow behaviour. It would also be valuable to investigate whether these behaviours are consistent across CsI:Tl detectors from different manufacturers and detector models. This validated model can be used to predict and compensate for detector lag during image acquisition, to improve the accuracy of quantitative x-ray imaging.

\section{Conclusions}
\label{sec:conclusions}
This work presented a comprehensive framework for investigating detector lag in flat-panel x-ray detectors with inorganic CsI:Tl scintillators, combining a reproducible experimental methodology with rate equation-based modelling. Detector lag measurement revealed repeatable overshoot behaviour on the $\sim$12-hour timescale during step-up transitions, together with other behaviours that are not predicted by existing detector lag models. These experimental observations were successfully characterised using a multi-exponential fitting approach, providing consistent estimates of the detector lag rates and coefficients. Building on these results, we proposed modifications to the existing multiple-trap rate equation model by introducing a nonlinear recombination term, non-radiative recombination pathways, and a generalised recombination exponent. These modifications successfully reproduced the key trends in the experimentally observed detector response, particularly the overshoot behaviour and the scaling that is absent from previous models.

Although the proposed model does not yet capture all aspects of the measured lag behaviour, it provides improved insight into the mechanisms governing detector lag and establishes a foundation for future model development. The experimental methodology and analysis framework developed in this work should also provide a reliable basis for future studies aimed at understanding and ultimately mitigating detector lag in x-ray imaging systems.

\section*{Acknowledgements}
The authors would like to sincerely thank Dr.~Lindsey Bignell for introducing the rate-equation framework used in this work and for many insightful discussions on the underlying detector physics. The authors also gratefully acknowledge Dr.~Lindon Roberts for his guidance on optimisation methods used to fit the rate-equation model to the experimental data. Yiyue Huang is funded through a Research Training Program (RTP) international stipend scholarship from the Australian Government.

\appendix
\section{Appendix}
\label{Appendix}
\subsection{Steady State solutions of rate equations}
\label{sec:appendix_steady_state}
\noindent
Here we derive steady-state trap populations, using Eqs.~\ref{eq.proposed1.1}--\ref{eq.proposedI} with $A_{nr,i}=0$, i.e., excluding the nonlinear depletion term (green). We also operate in the unsaturated regime where $n_i \ll N_i$ and $n_h \ll N_h$. Aside from short-lived transients, electrons can be considered to pass straight through the conduction band, with the great majority undergoing radiative recombination. Following \cite{bartram2006suppression}, we therefore set $\frac{d n_c}{d t} \approx 0$ in Eq.~\ref{eq.proposed1.1}, giving:
\[
  n_c \approx \frac{f_r + \sum_j q_j n_j}{\sum_j N_j A_{e,j} + A_{r}\,m^\beta}.
\]

\noindent Therefore, $n_c$ can be eliminated from Eq.~\ref{eq.proposed1.2}, which becomes:
\[
\frac{d n_i}{dt} = \frac{f_r + \sum_j q_j n_j}{\sum_j N_j A_{e,j} + A_{r}n_h^{\color{darkred}\beta}} N_i A_{e,i} - (q_i + {\color{darkred}A_{nr,i} n_h }) n_i.
\]
The emitted scintillation intensity can now be written in terms of the total carrier creation rate, minus the total rate of increase in trap population and the rate of non-radiative recombination:
\[ I = f - \sum_i \left(\frac{dn_i}{dt} + {\color{darkred}A_{nr,i}\, n_h \,n_i} \right).\]
Similarly, free holes in the valence band will be short-lived, so we can set $\dfrac{d m_v}{d t} \approx 0$ in Eqs.~\ref{eq.proposed1.3} and \ref{eq.proposed1.4} to eliminate $m_v$. We can also eliminate $n_h$ using $n_h = \sum_i n_i$ (the total numbers of electrons and holes are equal), leaving $n_i$ as the only significant independent variables.

For convenience, define
\[
\delta_r = \frac{A_r}{\sum_j N_j A_{e,j}} ,\qquad 
\delta_i=\frac{N_i A_{e,i}}{\sum_j N_j A_{e,j}},
\]
where $\sum_i \delta_i = 1$.  This gives:
\begin{align}
\frac{d  n_i}{dt}  &= \frac{ f_r + \sum_j q_j  n_j}{1 + \delta_r \left(\sum_j n_j\right)^{\color{darkred}\beta}}  \; \delta_i - \left(q_i + {\color{darkred}  A_{nr,i} \sum_j {n}_j}  \right) n_i, \label{eq:bart2}
\end{align}

\noindent
We now find the steady-state solutions $n_{0,i}$ for Eq.~\ref{eq:bart2}, with $A_{nr,i}=0$ (i.e., ignoring non-radiative recombination), by setting $\frac{dn_{0,i}}{dt} = 0$:
\begin{align} 
q_i {n}_{0,i} &= \delta_i \left( \frac{f + \sum_j q_j {n}_{0,j}}{1 + \delta_r\left(\sum_j n_{0,j}\right)^\beta } \right). \label{eq:steady1}
\end{align}
Since the right-hand term in Eq.~\ref{eq:steady1} is independent of $i$ it can be cancelled out allowing any trap occupancy to be eliminated in favour of any other one:
\[  {n}_{0,i} = \frac{\delta_i}{\delta_j} \frac{q_j}{q_i} {n}_{0,j}.\]
Using $\sum_j \delta_j = 1$ leads to the further relations:
\begin{align*}
\sum_j q_j {n}_{0,j} =
\frac{q_i}{\delta_i} {n}_{0,i} \quad (\text{for any }i), \\
\sum_j {n}_{0,j} =
\sum_j \frac{\delta_j}{q_j}\, \frac{q_i}{\delta_i}\, {n}_{0,i} = \bar{t}_t \, \frac{q_i}{\delta_i}\,{n}_{0,i},
\end{align*}
where $\bar{t}_t$ is the weighted mean electron trapping timescale $\bar{t}_t = \sum_j \frac{\delta_j}{q_h}$.   

Relating the trap occupancies $n_{0,i}$ to each other enables us to decouple the equations \ref{eq:steady1}:
\[
\frac{q_i}{\delta_i} {n}_{0,i} = \frac{{f} + \frac{q_i}{\delta_i} {n}_{0,i}}{1 +  \delta_r\left(\bar{t}_t \, \frac{q_i}{\delta_i}\,{n}_{0,i}\right)^\beta} \; , \label{eq:steady2} 
\]
giving the steady state solution to Eq.~\ref{eq:bart2}
\begin{align}
{n}_{0,i} &= \frac{\delta_i}{q_i} \, \left(\frac{{f}}{\delta_r \,\bar{t}_t^{\,\beta}}\right)^{\!\frac{1}{1+\beta}}.\label{eq:steady3}
\end{align}

This shows the scaling relation that $n_{o,i} \propto f^{1/1+\beta}$ - i.e. a smaller fraction of the incoming photons end up in traps as the intensity increases.  This is due to the nonlinear term $A_r n_c m^\beta$ in Eq.~\ref{eq.moretti1.1} that scintillates near-instantly, bypassing the traps. Conversely, the traps -- and therefore the amount of delayed scintillation -- play a more significant role at lower values of the intensity, $f$.

\subsection{Experimental Evidence for $\beta \sim 0.5$}
\label{sec:appendix_beta_evidence}
We now show that the data from \cite{starman2012nonlinear} and from our experiments are consistent with a $\beta$ value close to 0.5, rather than the value of 1 assumed in previous models. A non-integer value of $\beta$ is consistent with fractional-order kinetics, although to our knowledge, this has not previously been noted for CsI:Tl.

\begin{figure}
\centering
\includegraphics[width=.75\linewidth]{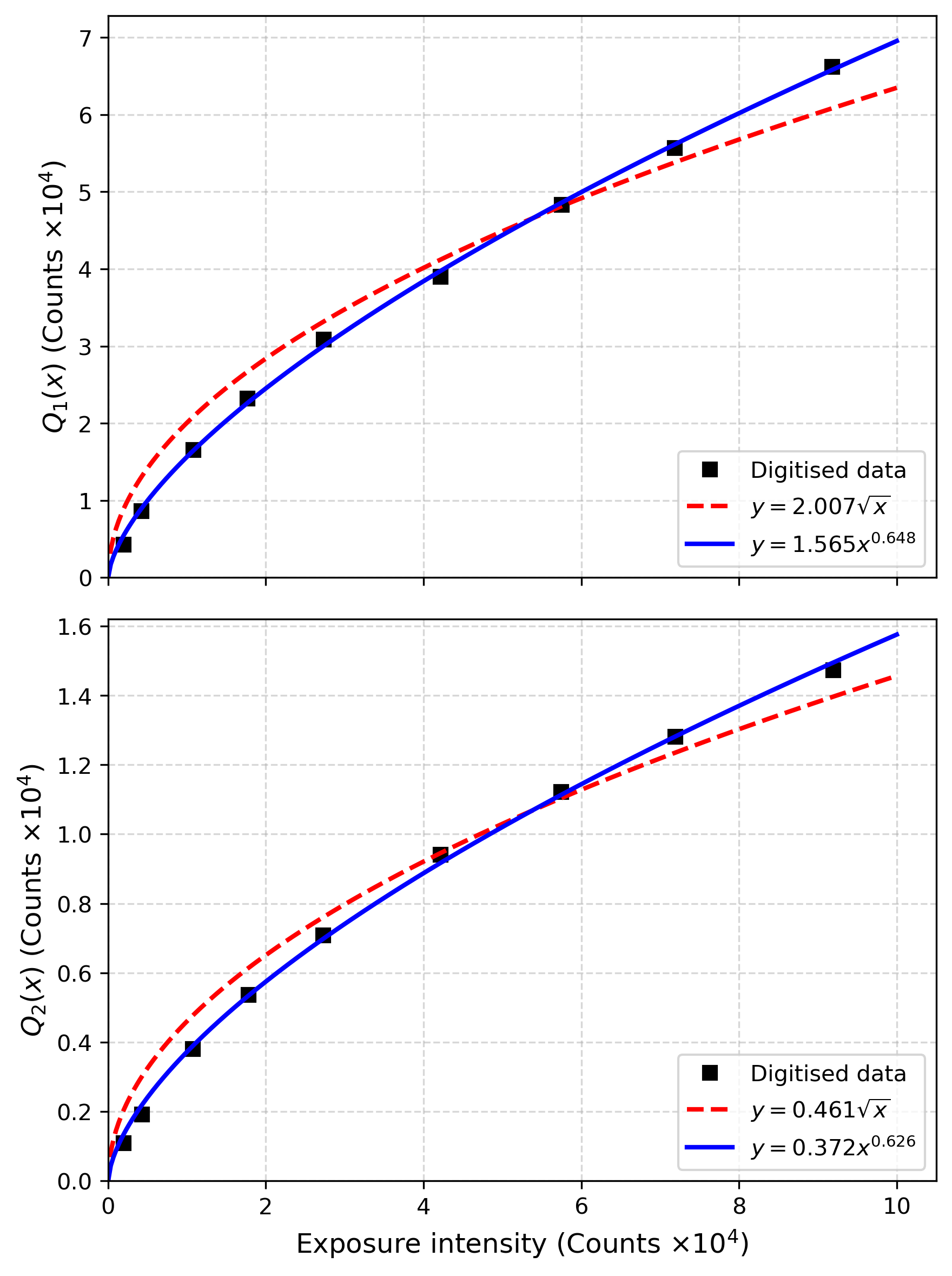}
\caption{Stored charge estimates, $Q_1(x)$ and $Q_2(x)$, digitised from \cite{starman2012nonlinear}. The data are fitted with both a square-root model ($Q\propto x^{0.5}$) and a general power-law model ($Q\propto x^{a}$). The fitted power-law exponents are 0.648 and 0.626, which are both close to $2/3$. These values are consistent with $\beta\sim0.5$ proposed in this work.}
\label{fig:starmanQn}
\end{figure}

\begin{figure}
\centering
\includegraphics[width=.9\linewidth]{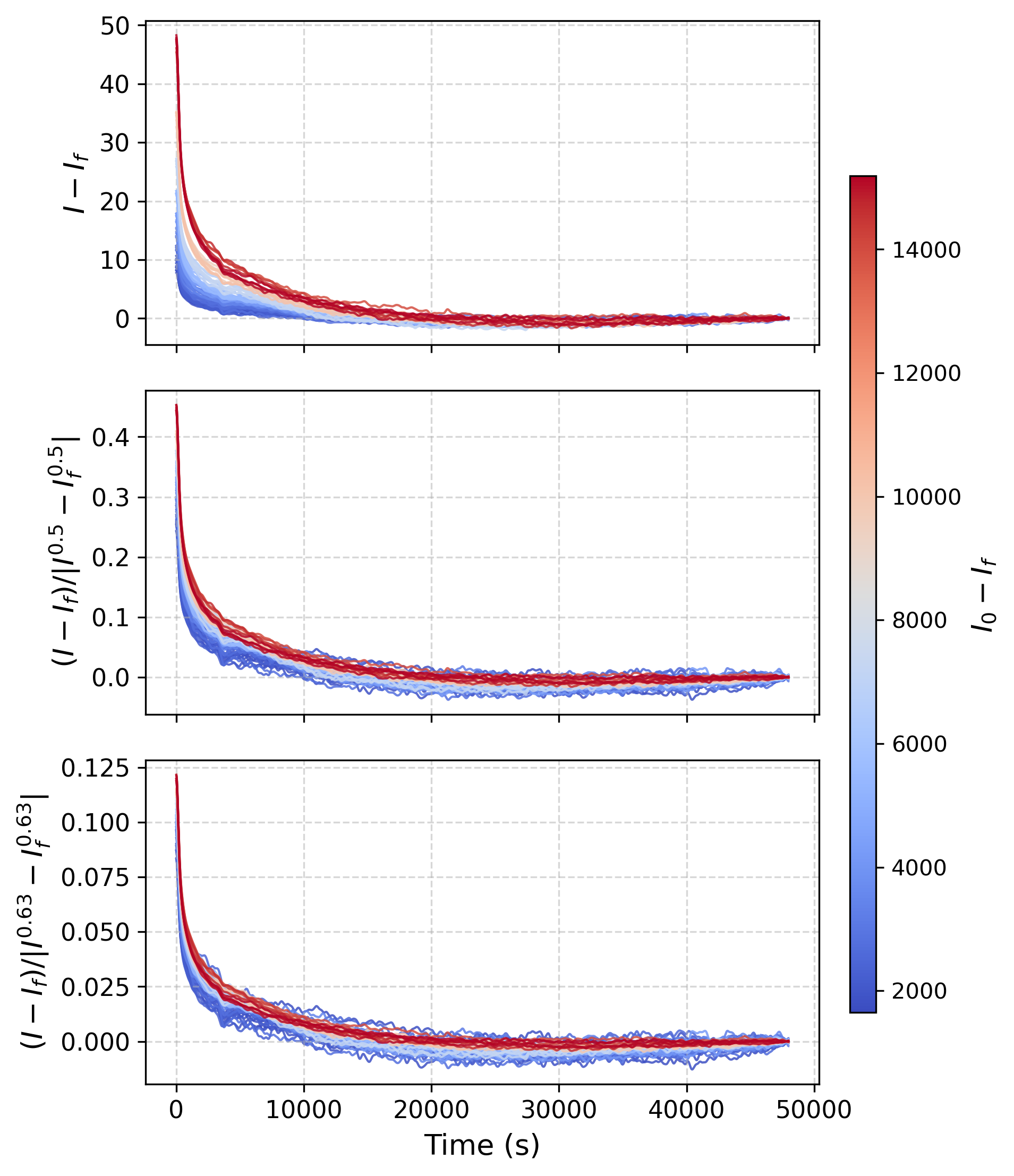}
\caption{Cu2$\rightarrow$DF1 (afterglow) transition curves scaled in different ways: top: unscaled (the same as presented in Fig.~\ref{fig:intensityplots}); middle: scaled with $\beta=1$; bottom: scaled with $\beta\sim0.5$. The bottom panel, with $\beta\sim0.5$, provides the best overlap of the transition curves over a range of intensity changes.}
\label{fig:afterglowcomparsion}
\end{figure}

The stored charge estimates, $Q_n$, reported in Figs.~\ref{fig:starmanQn}(a) and \ref{fig:starmanQn}(b) of \cite{starman2012nonlinear} were fitted using both a square-root model and a general power-law model. The fitted power-law exponents are 0.648 and 0.626 for $Q_1(x)$ and $Q_2(x)$, respectively, both of which are close to $2/3$. An exponent of approximately $2/3$ corresponds to $\beta=0.5$. The power-law scaling was not recognised in \cite{starman2012nonlinear} because their goal was to obtain an empirical fit, which was achieved using a fourth-order polynomial.

Figure~\ref{fig:afterglowcomparsion} shows the Cu2$\rightarrow$DF1 (afterglow) transition using different scaling values of $\beta$. The results again show that $\frac{1}{1+\beta} \sim 0.63$, or $\beta \sim 0.58$, provides the best overlap of the transition curves over a range of intensity changes. Only the afterglow transition is considered here, rather than transitions between different step wedges, because the rising baseline observed during intensity transitions does not reach equilibrium within the measurement period.

\subsection{Model optimisation and parameters}
\label{sec:appendix_model_optimisation}

For the optimisation, each fitted parameter was allowed to vary within bounds of 0.01 and 100 times its initial estimate, giving a range spanning four orders of magnitude. The exception was $\beta$, which was allowed to vary from 0.25 to 1.0. The logarithms of the parameters were taken before being passed to the DIRECT optimiser, to compress the allowed ranges and provide better uniformity between parameters. The initial estimates were first taken from \cite{moretti2016deep} and from the time constants obtained from the multi-exponential fits. Where an optimised parameter approached the edge of its prescribed bounds, its initial estimate was subsequently adjusted and the optimisation repeated.

\begin{table}[ht]
    \centering
    \caption{Optimised parameters used in the proposed rate-equation model. Parameters marked as fixed were not included in the optimisation. Where applicable, the corresponding values reported by Moretti et al.~\cite{moretti2016deep} or obtained from the multi-exponential fits are indicated in the Notes column.}
    \label{tab:model_parameters}
    \begin{tabular}{lll}
        \hline
        \textbf{Parameter (units)} & \textbf{Optimised value} & \textbf{Notes} \\
        \hline
        $\alpha$ & 0.7 & Fixed \\
        $f$ (s$^{-1}$ cm$^{-3}$) & $3.0\times10^{12}$ & Fixed \\
        $A_{e1}$ (cm$^{3}$ s$^{-1}$) & $3.0\times10^{-16}$ & Moretti: $1.0\times10^{-14}$ \\
        $A_{e2}$ (cm$^{3}$ s$^{-1}$) & $4.0\times10^{-16}$ & Moretti: $1.0\times10^{-15}$ \\
        $A_{e3}$ (cm$^{3}$ s$^{-1}$) & $1.0\times10^{-16}$ & \\
        $A_{e4}$ (cm$^{3}$ s$^{-1}$) & $3.6\times10^{-16}$ & \\
        $A_{e5}$ (cm$^{3}$ s$^{-1}$) & $1.0\times10^{-16}$ & \\
        $N_i$ ($i=1,\ldots,5$) (cm$^{-3}$) & $5.0\times10^{18}$ & Assumed identical \\
        $q_1$ (s$^{-1}$) & $8.0\times10^{-3}$ & Exponential fit: $1/\tau_1$ \\
        $q_2$ (s$^{-1}$) & $9.0\times10^{-4}$ & Exponential fit: $1/\tau_2$ \\
        $q_3$ (s$^{-1}$) & $2.33\times10^{-4}$ & Exponential fit: $1/\tau_3$ \\
        $q_4$ (s$^{-1}$) & $4.0\times10^{-7}$ & \\
        $q_5$ (s$^{-1}$) & $2.0\times10^{-8}$ & \\
        $M$ (cm$^{-3}$) & $1.0\times10^{19}$ & Moretti: $1.0\times10^{19}$ \\
        $A_r$ (cm$^{3}$ s$^{-1}$) & $1.0\times10^{-6}$ & Moretti: $1.0\times10^{-8}$ \\
        $A_h$ (cm$^{3}$ s$^{-1}$) & $1.0\times10^{-7}$ & Moretti: $1.0\times10^{-7}$ \\
        $q_{\mathrm{nr},4}$ (s$^{-1}$) & $4.0\times10^{-10}$ & \\
        $q_{\mathrm{nr},5}$ (s$^{-1}$) & $2.0\times10^{-5}$ & \\
        $A_{\mathrm{nr}}$ (cm$^{6}$ s$^{-1}$) & $1.0\times10^{-31}$ & \\
        $\beta$ & 0.5 & \\
        \hline
    \end{tabular}
\end{table}

The trap densities, $N_i$, were assumed to be identical for all five traps. This is considered reasonable because only the longest-lived traps have a chance of reaching saturation, as the trap population is approximately inversely proportional to $q_i$. In addition, the different $A_{e,i}$ values allow the traps to have different filling rates despite having identical trap densities.

\bibliographystyle{cas-model2-names}

\bibliography{refs}


\end{document}